\documentclass[article]{revtex4}

\usepackage{helvet}

\usepackage[UKenglish]{babel}
\usepackage{amsmath}
\usepackage{amssymb}
\usepackage{fancyhdr}
\usepackage{natbib}
\usepackage{xcolor}
\usepackage{ae,aecompl}
\usepackage{graphicx}
\usepackage{palatino}
\usepackage[T1]{fontenc}
\usepackage{rotating}
\usepackage{epsf}
\usepackage{setspace}
\usepackage{longtable}
\usepackage{wrapfig}
\usepackage{eurosym}
\usepackage{textcomp}
\usepackage{tabularx}
\usepackage{csquotes}

\usepackage{hyperref}

\usepackage{epstopdf}% To incorporate .eps illustrations using PDFLaTeX, etc.

\setcitestyle{sort}

\begin{document}

\title{Theory of groove-envelope phase effects in self-diffraction}

\author{Jan Reislöhner, 
Christoph G. Leithold
and Adrian N. Pfeiffer}

\affiliation{Institute of Optics and Quantum Electronics, Abbe Center of Photonics, 
Friedrich Schiller University, Max-Wien-Platz 1, 07743 Jena, Germany}

\begin{abstract}
\hspace{-0.4cm}
If two laser beams cross in a medium under shallow angle, the laser-induced grating consists of only a few grooves. In this situation, the phase between the grooves of the grating and its envelope is a decisive parameter for nonlinear effects. 
Here, models are established for reproducing the groove-envelope phase effects that have been observed in the interference pattern of self-diffraction. 
Four-wave mixing leads to interferences that are dominant in the spatial region between the orders of diffraction and with tilted interference fringes in the diagram of transverse coordinate vs. pulse delay. 
The vertical interference fringes that are dominant directly on the diffraction orders, experimentally observed at high intensity close to the damage threshold, require a model beyond four-wave mixing. 
A model is suggested that is based on optical transmission changes with confinement to regions in the medium that are smaller than the groove spacing. 
\end{abstract}

\maketitle

\parindent5mm

\section{Introduction}
\hspace{-0.6cm}
Strong laser fields in transparent solids change the refractive index. Many of the related phenomena can be described by the optical Kerr effect \cite{RN1027}, which is based on the third-order nonlinear response of the medium \cite{RN1049}. 
Since recently, there is a rapidly growing interest in situations where transparent solids are exposed to very intense few-cycle laser pulses, close to the damage threshold of the material. An important motivation is that high-order harmonic generation has been realized in transparent solids (see for example \cite{RN982, ISI:000331162400014, ISI:000356782900046, RN981, ISI:000378270300047}). The up-conversion of laser pulses to high-harmonic frequencies gains new opportunities because of the high density as compared to gas-phase targets. 
Another motivation stems from the potential of spectroscopic methods that are applicable to wide-bandgap dielectrics \cite{ISI:000312933800033, ISI:000376007200045, ISI:000376962300038}, giving new insight into the ultrafast dynamics of correlated many-electron systems. Furthermore, there is even the prospect of finding new schemes for ultrafast optoelectronics \cite{ISI:000312933800032}. 

For the accurate interpretation of spectroscopic methods as well as for the efficient exploitation of high-order harmonic generation from solid targets, effects of pulse propagation in macroscopic media under conditions of extreme nonlinearity need to be understood. Not only simple configurations with single beams are of interest, but also multi-beam configurations and beams with spatial-temporal coupling are relevant. Applications include the use of wavefront-rotation, which can be used for isolating attosecond pulses and for ultrafast spectroscopy \cite{ISI:000337721200005}. Governed by similar principles, two pulses in close-to-collinear configuration can be used for gating of the high-order harmonic generation \cite{ISI:000338942400001, RN979, ISI:000364483200010}. The regime of close-to-collinear configurations is barely explored, but it holds a number of interesting phenomena and opportunities. At the intersection of the beams, a close-to-collinear configuration causes an intensity grating that consists of only a few grooves. The groove-envelope offset phase (GEP), defined as the phase between the grooves and the spatial envelope of the beams, can be tuned by varying the delay between the pulses, where a delay of one optical cycle translates into a GEP shift of 2$\pi$. For close-to-collinear configurations, phenomena of nonlinear optics, such as probe retardation \cite{pati:2015,reislohner:2016} or self-diffraction \cite{pati:2016}, yield sub-cycle dependent signals with imprints of ultrafast strong-field processes.

In a non-collinear beam configuration, the two beams can be diffracted on the grating they form if the medium response is nonlinear, a process termed self-diffraction \cite{RN1030,RN1051}. The self-diffraction orders can be used for pulse characterization \cite{RN1048} as well as for spectroscopy of the medium itself \cite{ISI:A1975AG57700013}. Recently, it has been demonstrated that self-diffraction can be subcycle dependent (or equivalently GEP dependent) in a close-to-collinear configuration \cite{pati:2016}. Characteristic interference pattern have been observed. The first kind of interference pattern is dominant in the region between the diffraction orders and shows tilted interference fringes in the diagram of transverse coordinate $x$ vs. pulse delay $\tau$. The second kind of interference pattern, observed only at high intensities close to the damage threshold, is dominant on the diffraction orders and shows vertical interference fringes. 

As will be shown in this paper, the first kind of interference pattern can be regarded as a spatial analogue to the signal of an $f$-2$f$ interferometer that depends on the carrier-envelope offset phase (CEP) for few-cycle pulses \cite{RN965,RN974,RN977}. The CEP is a decisive parameter for nonlinear processes in the few-cycle regime \cite{RN960, RN985, RN959}, and likewise the GEP is a decisive parameter for nonlinear processes in the few-groove regime. The interferences are observed in the spectral overlap between a fundamental in a frequency-doubled pulse, and likewise the first kind of interference pattern is observed in the spatial overlap between spatial harmonics (diffraction orders) of the beams. In this respect, the observation of the second kind of interference pattern was very surprising, because the corresponding situation for an $f$-2$f$ interferometer would be that the spectrum becomes CEP dependent directly on the fundamental and the frequency-doubled wavelengths. 

The aim of this paper is to give a detailed treatment of analytical models describing the interferences in between and on the self-diffraction orders and to distinguish their contributions. Four-wave mixing (FWM), covered in section \ref{sec:fwm}, yields an interference pattern of the first kind. Evenly-spaced filamentation (ESF), covered in section \ref{sec:esf}, also yields an interference pattern of the first kind. Localized transmission changes (LTC), covered in section \ref{sec:ltc}, is the only mechanism studied here that yields an interference pattern of the second kind. Unlike most treatments, all the dependencies on the phases, both CEP and GEP, are explicitly taken into account.

\section{Definition of the laser induced grating}

\begin{figure*}[h]
\includegraphics[width=13cm]{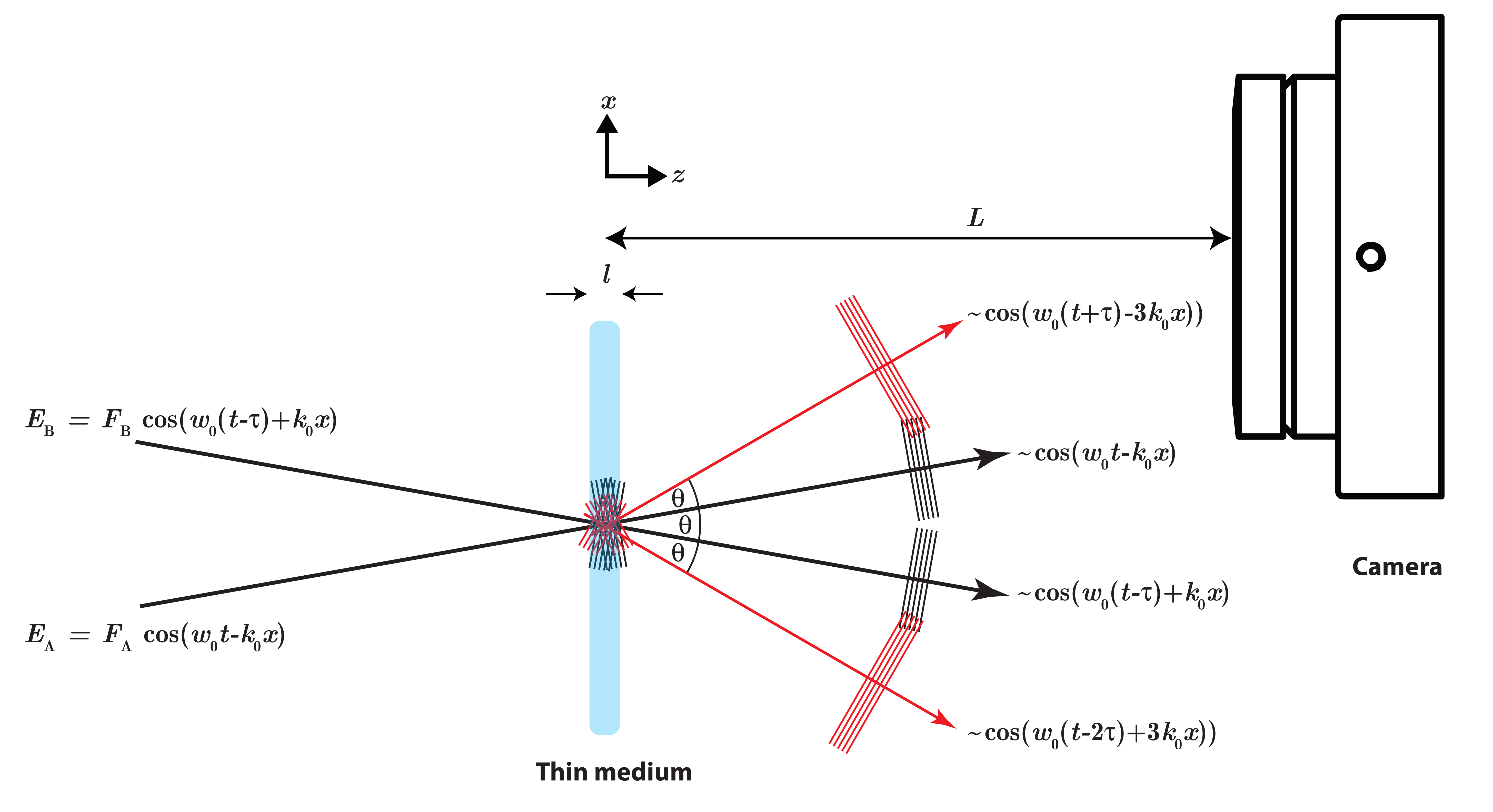}
\caption{\label{fig:SETUP} Laser pulses A and B interact in a nonlinear medium with variable pulse delay $\tau$. The thin medium has an infinitesimal thickness $l$. The light is detected after the macroscopic distance $L$ by a camera, which is positioned in $x-$direction such that the first diffraction order next to laser pulse A and the edge of laser pulse A are detected. The main part of laser pulse A is blocked in the experiment \cite{pati:2016}, because its intensity is much higher then the diffraction intensity.}
\end{figure*}

Laser pulses A and B with electric fields 
\begin{align}
E_\mathrm{A}(t,x) = F_\mathrm{A} \mathrm{cos}(\omega_0 t - k_0 x + \varphi_\mathrm{CEP})
\nonumber\\
E_\mathrm{B}(t,x) = F_\mathrm{B} \mathrm{cos}(\omega_0 (t-\tau) + k_0 x + \varphi_\mathrm{CEP})
\end{align}

interact in a nonlinear medium at $z = 0$ with variable pulse delay $\tau$ (see Figure \ref{fig:SETUP}). 
The CEP $\varphi_\mathrm{CEP}$ determines the temporal position of the carrier wave underneath the temporal envelope. In the experiment reported in Ref. \cite{pati:2016}, the CEP is not stabilized from shot-to-shot, but is identical for A and B in each shot (interferometric stability). The GEP $\varphi_\mathrm{GEP}$, defined by
\begin{equation}
\varphi_\mathrm{GEP} =  \omega_0  \tau,
\end{equation}
is adjusted by the pulse delay $\tau$. The GEP is the phase between the grooves of the intensity grating, which forms at $z = 0$ (see equation (\ref{eq:IntGrating})), and the spatial envelope of the beams.

The pulse envelopes
\begin{align}
F_\mathrm{A} = A_0 F(t,x) 
\nonumber\\
F_\mathrm{B} = B_0 F(t-\tau,x) 
\end{align}

are Gaussian functions in both temporal and spatial dimensions defined by

\begin{equation}
F(t,x) = e^{-\frac{1}{2\sigma} t^2} e^{-\frac{1}{2\kappa} x^2}.
\end{equation}

$A_0$ and $B_0$ are the amplitudes; $\sigma$ and $\kappa$ determine the pulse duration and the focal spot size, respectively. 

It should be noted that within this definition of the fields, their propagation direction depends on the wavelength, but in a usual experimental configuration all wavelength components of a pulse propagate into the same direction. It has been checked for all calculations presented in this paper that this approximation does not cause any significant deviation from a numerical calculation where all the wavelength components propagate into the same direction. The analytical form of the laser grating used here greatly simplifies the following analytical calculations. 

\section{\label{sec:fwm} Four-wave mixing}

\subsection{Nonlinear response}

In case of third-order nonlinearity and instantaneous response, the nonlinear polarization response of the medium is (atomic units are used throughout the paper unless otherwise stated)
\begin{align}
P^{\mathrm{NL}}(t,x) = \chi^{(3)} \left( E_\mathrm{A}(t,x) + E_\mathrm{B}(t,x) \right)^3 
&= \chi^{(3)} ( \frac{3F_\mathrm{A}^3 + 6 F_\mathrm{A} F_\mathrm{B}^2 }{4}  \mathrm{cos}(\omega_0 t - k_0 x + \varphi_\mathrm{CEP})
\nonumber\\
&+ \frac{3F_\mathrm{B}^3 + 6 F_\mathrm{A}^2 F_\mathrm{B} }{4}  \mathrm{cos}(\omega_0 t + k_0 x - \varphi_\mathrm{GEP} + \varphi_\mathrm{CEP})
\nonumber\\
&+ \frac{3 F_\mathrm{A}^2 F_\mathrm{B} }{4}  \mathrm{cos}(\omega_0 t - 3 k_0 x + \varphi_\mathrm{GEP}  + \varphi_\mathrm{CEP})
\nonumber\\
&+ \frac{3 F_\mathrm{A} F_\mathrm{B}^2 }{4}  \mathrm{cos}(\omega_0 t + 3 k_0 x - 2 \varphi_\mathrm{GEP} + \varphi_\mathrm{CEP})
\nonumber\\
&+ \frac{F_\mathrm{A}^3 }{4}  \mathrm{cos}(3\omega_0 t - 3 k_0 x + 3 \varphi_\mathrm{CEP})
\nonumber\\
&+ \frac{F_\mathrm{B}^3 }{4}  \mathrm{cos}(3\omega_0 t + 3 k_0 x - 3 \varphi_\mathrm{GEP} + 3 \varphi_\mathrm{CEP})
\nonumber\\
&+ \frac{3 F_\mathrm{A}^2 F_\mathrm{B} }{4}  \mathrm{cos}(3 \omega_0 t - k_0 x - \varphi_\mathrm{GEP}  + 3 \varphi_\mathrm{CEP})
\nonumber\\
&+ \frac{3 F_\mathrm{A} F_\mathrm{B}^2 }{4}  \mathrm{cos}(3 \omega_0 t + k_0 x - 2 \varphi_\mathrm{GEP}  + 3 \varphi_\mathrm{CEP}) ).
\end{align}

All terms oscillating at angular frequency $\omega_0$ include the phase $\varphi_\mathrm{CEP}$, whereas all terms oscillating at angular frequency $3 \omega_0$ include the phase $3 \varphi_\mathrm{CEP}$. If interference occurs between light at the fundamental frequency and light at the third harmonic frequency, then the detected intensity will be dependent on the CEP. However, interference between the fundamental and the third harmonic would require an extremely large bandwidth. In case of a second-order nonlinearity in the medium, the interference between the fundamental and the second harmonic depends on the CEP likewise, and the required spectral bandwidth is reduced. This effect is exploited in a so-called $f$-2$f$ interferometer to detect CEP drifts \cite{RN965,RN974,RN977}. 

In the following, the terms oscillating at angular frequency $3\omega_0$ are neglected, because light at the third harmonic frequency is absorbed in the medium used in Ref. \cite{pati:2016}. All terms oscillating at angular frequency $\omega_0$ include the phase $\varphi_\mathrm{CEP}$, as laser pulses A and B do, so the intensity that is detected after the medium does not depend on the CEP. Therefore, to simplify the following equations, the CEP dependence is not explicitly included in the following analysis by setting 
\begin{equation}
\varphi_\mathrm{CEP} = 0. 
\end{equation}

The detection with a camera is restricted to the spatial region of laser pulse A and the neighboring diffraction order (see Figure \ref{fig:SETUP}). For the laser parameters of the analyzed experiment \cite{pati:2016}, only the terms with phase dependencies $- k_0 x$ and $- 3 k_0 x$ need to be accounted for, whereas the terms with phase dependencies $k_0 x$ and $3 k_0 x$ can be neglected. For experimental conditions with significantly smaller crossing angles between A and B or smaller focal waists, all terms could potentially influence the spatial region between laser pulse A and the neighboring diffraction order. This would lead to interferences that are GEP-dependent with phases $2 \varphi_\mathrm{GEP}$ and $3 \varphi_\mathrm{GEP}$, see \cite{reislohner:2016}. In the experiment described in the Ref. \cite{pati:2016}, all observed interferences dependent on the GEP with phase $\varphi_\mathrm{GEP}$; faster oscillations are not observed. Therefore, the effective nonlinear polarization response for the following analysis is 
\begin{equation}
\label{eq:PEFFECTIVE}
P^{\mathrm{EF}}(t,x) = \chi^{(3)} \left( P_{\mathrm{AAA}}(t,x) + P_{\mathrm{ABB}}(t,x)  + P_{\mathrm{AAB}}(t,x) \right)
\end{equation}

with the definitions
\begin{align}
P_{\mathrm{AAA}}(t,x) &= \frac{3F_\mathrm{A}^3 }{4}  \mathrm{cos}(\omega_0 t - k_0 x) 
\nonumber\\
P_{\mathrm{ABB}}(t,x) &= \frac{3 F_\mathrm{A} F_\mathrm{B}^2 }{2}  \mathrm{cos}(\omega_0 t - k_0 x)
\nonumber\\
P_{\mathrm{AAB}}(t,x)&= \frac{3 F_\mathrm{A}^2 F_\mathrm{B} }{4}  \mathrm{cos}(\omega_0 t - 3 k_0 x + \varphi_\mathrm{GEP} ).
\end{align}

The term $P_{\mathrm{AAA}}(t,x)$ is associated with self-phase modulation; the term $P_{\mathrm{ABB}}(t,x)$ is associated with cross-phase modulation; the term $P_{\mathrm{AAB}}(t,x)$ is associated with self-diffraction. The fact that the light from self-diffraction depends on the GEP, whereas the light from cross-phase modulation and self-phase modulation as well as pulse A do not depend on the GEP, leads to GEP-dependent interference (interference that depends on the pulse delay on a subcycle timescale). This can be regarded as the spatial analogue to an $f$-2$f$ interferometer: the CEP determines the interference with light at up-converted temporal frequencies, while the GEP determines the interference with light at up-converted spatial frequencies. 

\subsection{\label{sec:fwmPP} Pulse propagation}

For subsequent pulse propagation, the fields are Fourier transformed with the convention
\begin{align}
\tilde f(\omega,x) &\propto \int_{-\infty}^{+\infty}{ f(t,x)} e^{-i \omega t} dt
\nonumber\\
\hat f(\omega,k) &\propto   \int_{-\infty}^{+\infty}{ \tilde f(\omega,x)} e^{-i k x} dx.
\end{align}
The Fourier transform of laser pulse A is
\begin{align}
\hat E_\mathrm{A}(\omega,k) &\propto A_0 e^{-\frac{\sigma}{2} (\omega - \omega_0)^2} e^{-\frac{\kappa}{2} (k + k_0)^2}.
\end{align}
Similarly, the Fourier transform of the effective nonlinear polarization response is obtained:
\begin{align}
\hat P_{\mathrm{AAA}} (\omega,k)  &\propto \frac{1}{4} A_0^3 e^{-\frac{\sigma}{6} (\omega - \omega_0)^2} e^{-\frac{\kappa}{6} (k + k_0)^2}
\nonumber\\
\hat P_{\mathrm{ABB}} (\omega,k)  &\propto \frac{1}{2} A_0 B_0^2 e^{-\frac{\sigma}{6} (\omega - \omega_0)^2}  e^{-i \frac{2}{3} \tau (\omega - \omega_0)}  e^{-\frac{\kappa}{6} (k + k_0)^2} e^{- \frac{1}{3 \sigma} \tau^2}
\nonumber\\
\hat P_{\mathrm{AAB}} (\omega,k)  &\propto \frac{1}{4} A_0^2 B_0 e^{-\frac{\sigma}{6} (\omega - \omega_0)^2} e^{-i \frac{1}{3} \tau (\omega - \omega_0)}  e^{-\frac{\kappa}{6} (k + 3 k_0)^2}  e^{- \frac{1}{3 \sigma} \tau^2} e^{i \omega_0 \tau}.
\end{align}

In the limit of a thin medium with infinitesimal thickness $l$, the electric field after the medium is 

\begin{equation}
\label{sec:eqEl} 
\hat E^\mathrm{FWM}(\omega, k) = \hat E_\mathrm{A}(\omega,k) + \hat E_\mathrm{B}(\omega,k) -i \frac{2 \pi \omega l}{c}  \hat P^{\mathrm{NL}} (\omega,k),
\end{equation}
where $c$ is the speed of light. The linear polarization response is neglected here for simplicity, but it can be shown that the linear polarization response does not yield any additional interferences in self-diffraction. 

Restriction to the spatial region of laser pulse A and the neighboring diffraction order and restriction to the frequency $\omega= \omega_0$ yields
\begin{equation}
\label{eq:Ew0l}
\hat E^\mathrm{FWM}(\omega = \omega_0, k) \propto A_0 e^{-\frac{\kappa}{2} (k + k_0)^2}  + C_0 e^{-\frac{\kappa}{6}(k + k_0)^2}  +  D_0  e^{-\frac{\kappa}{6} (k + 3 k_0)^2} e^{i \omega_0 \tau}
\end{equation}

with the definitions
\begin{align}
C_0 &=  -i \frac{2 \pi \omega_0 l}{c}   \chi^{(3)} \left(  \frac{1}{4} A_0^3   +  \frac{1}{2} A_0 B_0^2   e^{- \frac{1}{3 \sigma} \tau^2}  \right)
\nonumber\\
D_0 &=  -i \frac{2 \pi \omega_0 l}{c}   \chi^{(3)} \left(  \frac{1}{4} A_0^2 B_0    e^{- \frac{1}{3 \sigma} \tau^2}  \right).
\end{align}

After the interaction with the thin nonlinear medium, the electric field propagates the macroscopic distance $L$ through free space to a camera. With the definition
\begin{equation}
\beta = \frac{1}{2} \frac{c}{\omega} L
\end{equation}
and within the paraxial approximation, the electric field at distance $L$ is given by

\begin{align}
\hat E^\mathrm{FWM}_L(\omega = \omega_0, k)  &\propto \hat E^\mathrm{FWM}(\omega = \omega_0, k)  e^{i \beta k^2}
\nonumber\\
&= \left( A_0 e^{-\frac{\kappa}{2} (k + k_0)^2}  + C_0 e^{-\frac{\kappa}{6}(k + k_0)^2}  +  D_0  e^{-\frac{\kappa}{6} (k + 3 k_0)^2} e^{i \omega_0 \tau} \right) e^{i \beta k^2}.
\nonumber\\
\end{align}

The inverse Fourier transform can be performed yielding a representation in space-domain:

\begin{align}
\label{eq:group}
\tilde E_L^\mathrm{FWM}(\omega = \omega_0, x)  \propto  A_L  e^{\lambda_A}  e^{i  \phi_A} + C_L e^{\lambda_C}  e^{i  \phi_C} +  D_L  e^{\lambda_D}  e^{i  \phi_D}
\end{align}

with the definitions

\begin{align}
 \lambda_A  &= -\frac{1}{1 + \frac{\beta^2}{(\kappa/2)^2}}  \left( \frac{1}{2 \kappa} (x - 2 k_0 \beta)^2 - \frac{2 k_0^2 \beta^2}{\kappa}  \right)
 \nonumber\\
\phi_A  &=  -\frac{1}{1 + \frac{\beta^2}{(\kappa/2)^2}}  \left( \frac{\beta}{\kappa^2} x^2 + k_0 x  \right) 
  \nonumber\\
\lambda_C  &= -\frac{1}{1 + \frac{\beta^2}{(\kappa/6)^2}}  \left( \frac{3}{2 \kappa} (x - 2 k_0 \beta)^2 - \frac{6 k_0^2 \beta^2}{\kappa} \right)
 \nonumber\\
\phi_C  &= -\frac{1}{1 + \frac{\beta^2}{(\kappa/6)^2}}  \left( \frac{9 \beta}{\kappa^2} x^2 + k_0 x \right)
  \nonumber\\
\lambda_D  &= -\frac{1}{1 + \frac{\beta^2}{(\kappa/6)^2}}   \left(  \frac{3}{2 \kappa} (x - 6 k_0 \beta)^2 - \frac{6 (3 k_0)^2 \beta^2}{\kappa}  \right)
 \nonumber\\
\phi_D  &=  \varphi_{GEP} -\frac{1}{1 + \frac{\beta^2}{(\kappa/6)^2}}  \left(  \frac{9 \beta}{\kappa^2} x^2 + 3 k_0 x \right) .
\end{align}

The terms with label $A$ originate from laser pulse A. The terms with label $C$ originate from self-phase modulation and cross-phase modulation. The terms with label $D$ originate from self-diffraction.

\subsection{Interference pattern}

The camera at position $z = L$ detects the intensity 

\begin{equation}
\label{eq:I}
I^\mathrm{FWM}(x, \varphi_\mathrm{GEP}) = \left| \tilde E_L^\mathrm{FWM}(\omega = \omega_0, x)   \right|^2.
\end{equation}

To analyze the interferences, the expression for the intensity is separated into terms that are independent respectively dependent on the GEP:
\begin{equation}
\label{eq:Ixtau}
I^\mathrm{FWM}(x, \varphi_\mathrm{GEP}) = I_0^\mathrm{FWM}(x) + I_{\mathrm{GEP}}^\mathrm{FWM}(x, \varphi_\mathrm{GEP}) 
\end{equation}

with 
\begin{align}
I_0^\mathrm{FWM}(x) &= \left| A_L  e^{\lambda_A}  \right|^2 + \left|  C_L e^{\lambda_C}  \right|^2 + \left| D_L  e^{\lambda_D}  \right|^2 
+ 2  \mathrm{Re}  \{  \left( A_L  e^{\lambda_A}  e^{i  \phi_A}   \right)  \left( C_L e^{\lambda_C}  e^{i  \phi_C} \right)^* \}
\end{align}

and
\begin{align}
\label{eq:Idefs}
I_{\mathrm{GEP}}^\mathrm{FWM}(x, \varphi_\mathrm{GEP}) &= 2 \mathrm{Re}  \{   I_{AD}  e^{i  \phi_{AD}}  \} + 2 \mathrm{Re}  \{   I_{CD}  e^{i  \phi_{CD}} \}
\nonumber\\
I_{AD} &= A_L  e^{\lambda_A}  D_L^*  (e^{\lambda_D})^*
\nonumber\\
I_{CD} &= C_L  e^{\lambda_C}  D_L^*  (e^{\lambda_D})^*
\nonumber\\
\phi_{AD} &= \phi_A - \phi_D = - \varphi_\mathrm{GEP}  -\frac{1}{1 + \frac{\beta^2}{(\kappa/2)^2}}  \left( \frac{\beta}{\kappa^2} x^2 + k_0 x  \right) +\frac{1}{1 + \frac{\beta^2}{(\kappa/6)^2}}  \left(  \frac{9 \beta}{\kappa^2} x^2 + 3 k_0 x \right) 
\nonumber\\
\phi_{CD} &= \phi_C - \phi_D = - \varphi_\mathrm{GEP} + \frac{1}{1 + \frac{\beta^2}{(\kappa/6)^2}}  \left( 2 k_0 x \right).
\end{align}

\begin{figure*}[h]
\includegraphics[width=13cm]{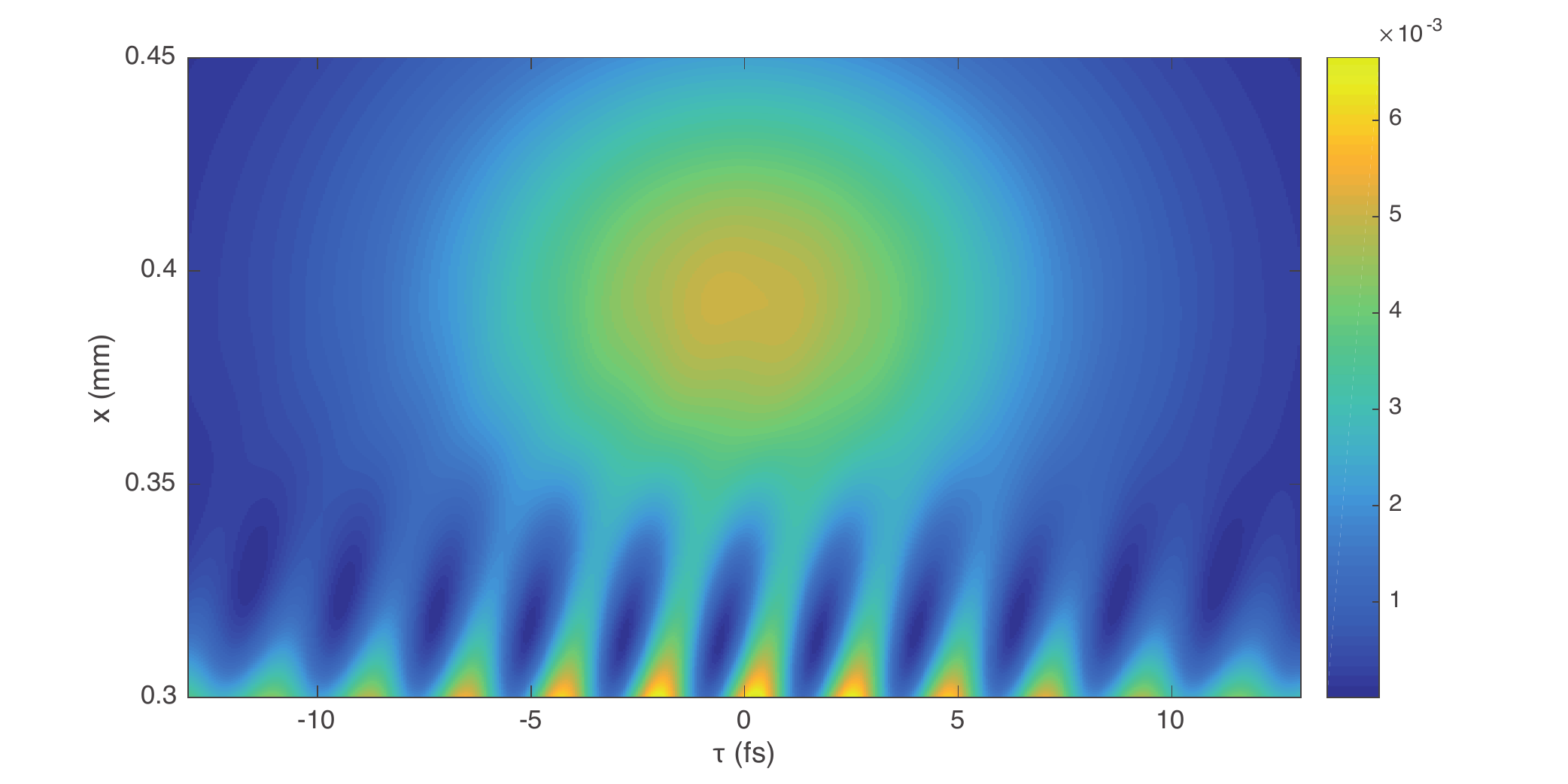}
\caption{\label{fig:FWM} Interferences in FWM according to equation (\ref{eq:I}). The displayed region of $x$ includes the edge of pulse A (at the bottom of the figure) and the center of the first diffraction order next to A, which is centered at about $x = 0.4$ mm. See Figure \ref{fig:SETUP} for the position of the camera. The parameters used are (in atomic units): $A_0 $ = 1,  $B_0$ = 0.1, $\omega_0$ = 0.067, $k_0$ = 4.27e-06, $\kappa$ = 8.75e+11, $\sigma$ = 6.16e+04, $\frac{2 \pi \omega_0 l}{c}   \chi^{(3)}$ = 1, $L =$ 2.83e+08. This corresponds to a crossing angle of 1$^\circ$, a beam waist of 70\,$\mu$m and a propagation distance of 15 mm. }
\end{figure*}

The expressions for $\phi_{AD}$ and $\phi_{CD}$ in equation (\ref{eq:Idefs}) reveal that the $\tau$-dependent interferences have a phase that depends on $x$. With the parameters of Figure \ref{fig:FWM}, $\frac{\beta}{\kappa^2}$ = 3.7e-13 and $k_0$ = 4.27e-06, so the phase depends essentially linearly on $x$ and can be approximated by
\begin{align}
\phi_{AD} &\approx  - \varphi_\mathrm{GEP} + \left( -\frac{1}{1 + \frac{\beta^2}{(\kappa/2)^2}}   +\frac{3 }{1 + \frac{\beta^2}{(\kappa/6)^2}}  \right) k_0 x 
\nonumber\\
\phi_{CD} &\approx - \varphi_\mathrm{GEP} + \frac{2 k_0}{1 + \frac{\beta^2}{(\kappa/6)^2}}  x .
\end{align}

The delay scans displayed in Figure \ref{fig:FWM} reveal oscillations with the periodicity of the optical cycle. The oscillations are most dominant in the region between pulse A and the first diffraction order, and they are tilted in the $x$-$\tau$-diagram. The first contribution to these oscillations is interference with phase $\phi_{AD}$ between the first diffraction order and pulse A. This contribution is dominant in the experiment reported in Ref. \cite{pati:2016}. Another contribution, which is weaker in the experiment, is interference with phase $\phi_{CD}$ between the the first diffraction order and light from self-phase modulation and cross-phase modulation. 

\begin{figure*}[h]
\includegraphics[width=13cm]{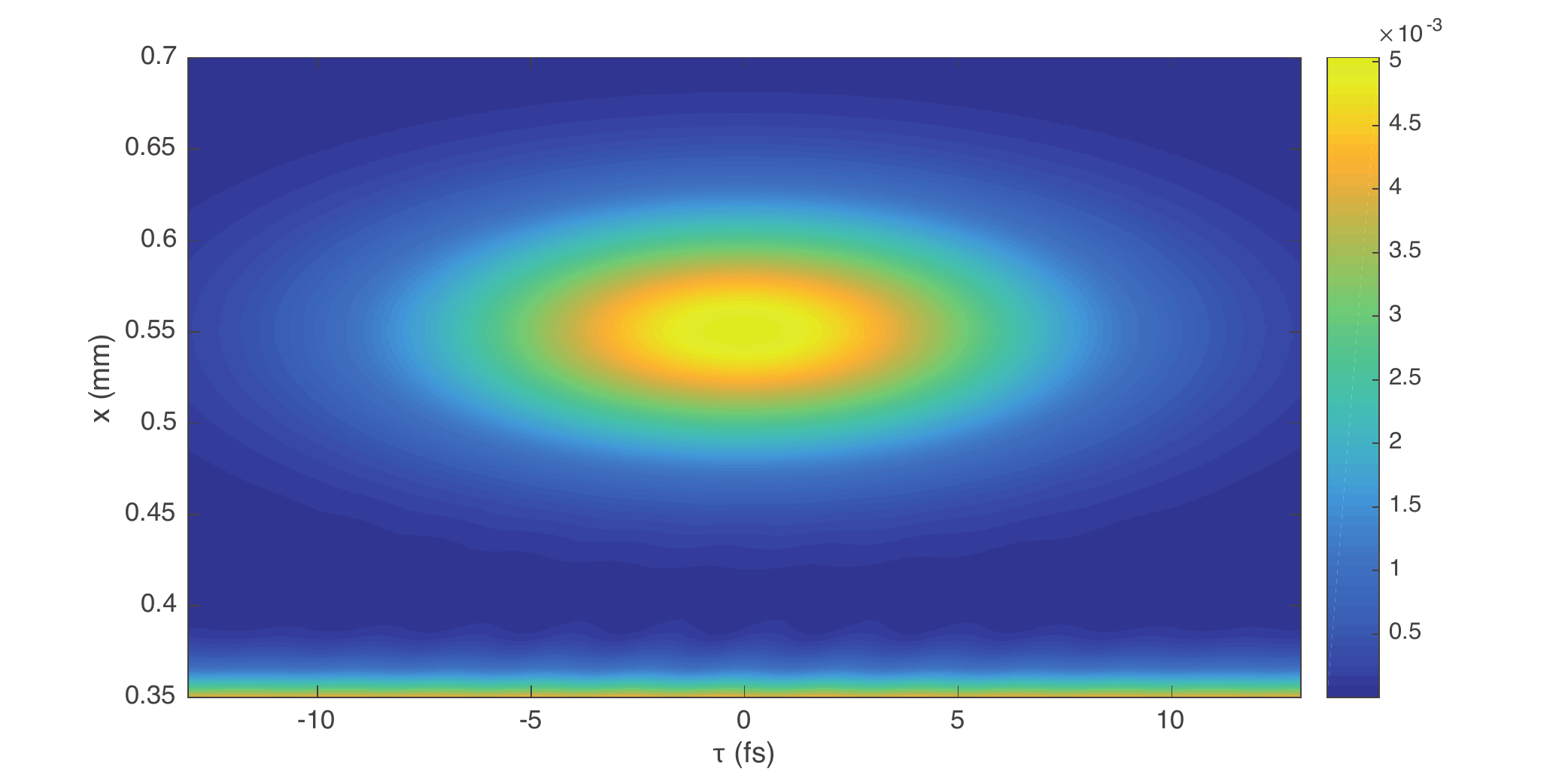}
\caption{\label{fig:FWM_many} Same as Figure \ref{fig:FWM}, except that the crossing angle is slighly increased to 1.4$^\circ$.}
\end{figure*}

The oscillations are visible only for few-groove gratings, which corresponds to small crossing angles and small beam waists. For larger crossing angles, the oscillations diminish very quickly (see Figure \ref{fig:FWM_many}). However, for comparison with an experiment where a medium with finite thickness is used, it needs to be considered that nonlinear defocusing in the sample widens the divergence angle, which enhances the contrast of the oscillations. 

\section{\label{sec:esf} Evenly-spaced filamentation}

\subsection{Nonlinear response}

When laser pulses A and B intersect in the nonlinear medium at position $z = 0$, the instantaneous intensity is 

\begin{align}
  \left( E_\mathrm{A}(t,x) + E_\mathrm{B}(t,x) \right)^2
&=  \frac{F_\mathrm{A}^2 }{2}  +  \frac{F_\mathrm{B}^2 }{2} + F_\mathrm{A} F_\mathrm{B} \mathrm{cos}(-2 k_0 x + \varphi_\mathrm{GEP} )
\nonumber\\
&+ \frac{F_\mathrm{A}^2 }{2}   \mathrm{cos}(2 \omega_0 t - 2 k_0 x + 2 \varphi_\mathrm{CEP})
\nonumber\\
&+  \frac{F_\mathrm{B}^2 }{2}  \mathrm{cos}(2 \omega_0 t + 2 k_0 x - 2 \varphi_\mathrm{GEP} + 2 \varphi_\mathrm{CEP})
\nonumber\\
&+ F_\mathrm{A} F_\mathrm{B} \mathrm{cos}(2 \omega_0 t  -  \varphi_\mathrm{GEP} + 2 \varphi_\mathrm{CEP}).
\end{align}

The time-average reveals the form of a grating,
\begin{align}
\label{eq:IntGrating}
\left<  \left( E_\mathrm{A}(t,x) + E_\mathrm{B}(t,x) \right)^2 \right>
&=  \frac{F_\mathrm{A}^2 }{2}  +  \frac{F_\mathrm{B}^2 }{2} + F_\mathrm{A} F_\mathrm{B} \mathrm{cos}(2 k_0 x - \varphi_\mathrm{GEP} ),
\end{align}
with maxima at positions $x_{max, n} = \frac{n \pi}{k_0} + \frac{\varphi_\mathrm{GEP} }{2 k_0}$.

The numerical calculation in Ref. \cite{pati:2016} shows that the grooves of the intensity grating are much steeper at the end of the medium compared to the beginning of the medium. This can be understood as the result of multiple filamentation, where the filaments do not build up at random positions due to wavefront distortions, but where the filaments are triggered by the grating at the beginning of the medium. The initial intensity distribution of the grating acts like a micro-lens array that initiates evenly-spaced filaments \cite{liu:2005}. Here it is important to note that the steepening of the filaments causes self-diffraction. The reason is that the steepening of the filaments broadens of the spatial frequencies, and the even spacing of the filaments corresponds to spatial harmonics (diffraction orders) in the far field. 

It may be argued that the two mechanisms for self-diffraction that are treated here, FWM and ESF, are not two distinct mechanisms, but interpretations of one effect, because both mechanisms originate from the nonlinear response of the medium. The conception of two separate mechanisms is nourished by two findings: i) self-diffraction can be treated by FWM only, without inclusion of filamentation. This case is treated in section \ref{sec:fwm}. It is also treated in Ref. \cite{pati:2016}, where the approximation of the numerical calculation (neglecting diffraction inside the sample) leads to self-diffraction without self-focusing. ii) ESF is very similar to diffraction on a real micro-lens array, like it is done in Ref. \cite{liu:2005}.

To investigate the self-diffraction that arises due to ESF with a simple model, it is assumed that the intensity distribution after the medium is given by
\begin{equation}
\label{eq:grating}
I(x) = I_0  e^{-\frac{1}{\kappa} x^2}  \sum_{q = 0}^{+\infty}{ a_q \mathrm{cos}(2 q k_o x - q  \varphi_\mathrm{GEP} ) } .
\end{equation}

The coefficients $a_q$ depend of course on the pulse delay $\tau$. At large pulse delays, the modulation depth of the intensity grating at the beginning of the medium is reduced. However, it is not straightforward to calculate explicitly the filamentation that follows, since this is a highly nonlinear process. Therefore the exact functional form of the coefficients $a_q$ will be omitted here. The GEP-dependence of the Fourier components in equation (\ref{eq:grating}) follows from the condition that the Fourier components should interfere constructively at positions $x_{max, n}$. 

Within the present model, it is assumed that the electric field after the medium is given by 
\begin{align}
\label{eq:eesf}
\tilde E^{\mathrm{ESF}}(\omega, x)  &=   \left( \tilde E_\mathrm{A}(\omega,x) + \tilde E_\mathrm{B}(\omega,x) \right) \sum_{q = 0}^{+\infty}{b_q \mathrm{cos}(2 q k_o x - q  \varphi_\mathrm{GEP} ) } .
\end{align}

The camera detects only the spatial region of laser pulse A and the neighboring diffraction order. Therefore, corresponding to the approximation made in equation (\ref{eq:PEFFECTIVE}), only the terms with phase dependencies $- k_0 x$ and $- 3 k_0 x$ need to be accounted for: 
\begin{align}
\tilde E^{\mathrm{ESF}}(\omega = \omega_0, x)  &= e^{-\frac{1}{2 \kappa}x^2} 
\left(   A_0 b_0 e^{-i k_0 x} + A_0 \frac{b_1}{2} e^{- 3 i  k_0 x } e^{i \omega_0 \tau}  + B_0 \frac{b_1}{2} e^{-i k_0 x} + B_0 \frac{b_2}{2} e^{-3 i k_0 x}  e^{i \omega_0 \tau}  \right). 
\end{align}

\subsection{Pulse propagation}

The Fourier transform of the field after the medium is
\begin{align}
\hat E^{\mathrm{ESF}}(\omega = \omega_0, k)  &\propto 
\left( \left( A_0 b_0 + B_0 \frac{b_1}{2} \right) e^{-\frac{\kappa}{2} (k + k_0)^2} 
+  \left( A_0  \frac{b_1}{2}  + B_0 \frac{b_2}{2} \right)  e^{-\frac{\kappa}{2} (k + 3 k_0)^2} e^{i \omega_0 \tau}   \right).
\end{align}

Pulse propagation by the distance $L$ and subsequent inverse Fourier transform can be performed similarly as in section \ref{sec:fwmPP}. The final expression is
\begin{align}
\tilde E_L^{\mathrm{ESF}}(x) \propto  K_L  e^{\lambda_K}  e^{i  \phi_K} + M_L e^{\lambda_M}  e^{i  \phi_M}
\end{align}

with the definitions

\begin{align}
K_L &= \left( A_0 b_0 + B_0 \frac{b_1}{2} \right) \sqrt{\frac{1}{\zeta_2}}  e^{- \frac{\kappa}{2} (1 - \frac{1}{\zeta_2}) k_0^2}
\nonumber\\
M_L &= \left( A_0  \frac{b_1}{2}  + B_0 \frac{b_2}{2} \right) \sqrt{\frac{1}{\zeta_2}}  e^{- \frac{\kappa}{2} (1 - \frac{1}{\zeta_2}) (3 k_0)^2}
\nonumber\\
 \lambda_K  &= -\frac{1}{1 + \frac{\beta^2}{(\kappa/2)^2}}  \left( \frac{1}{2 \kappa} (x - 2 k_0 \beta)^2 - \frac{2 k_0^2 \beta^2}{\kappa}  \right)
 \nonumber\\
\phi_K  &=  -\frac{1}{1 + \frac{\beta^2}{(\kappa/2)^2}}  \left( \frac{\beta}{\kappa^2} x^2 + k_0 x  \right) 
  \nonumber\\
\lambda_M  &= -\frac{1}{1 + \frac{\beta^2}{(\kappa/2)^2}}  \left( \frac{1}{2 \kappa} (x - 6 k_0 \beta)^2 - \frac{2 (3 k_0)^2 \beta^2}{\kappa}  \right)
 \nonumber\\
\phi_M  &= \varphi_\mathrm{GEP}  - \frac{1}{1 + \frac{\beta^2}{(\kappa/2)^2}}  \left( \frac{\beta}{\kappa^2} x^2 + 3 k_0 x  \right) .
\end{align}

\subsection{Interference pattern}

The camera at position $z = L$ detects the intensity 

\begin{equation}
\label{eq2:I}
I^\mathrm{ESF}(x, \varphi_\mathrm{GEP}) = \left| \tilde E_L^{\mathrm{ESF}}(x)  \right|^2 .
\end{equation}

To analyze the interferences, the expression for the intensity is separated into terms that are independent respectively dependent on the GEP:
\begin{equation}
\label{eq2:Ixtau}
I^\mathrm{ESF}(x, \varphi_\mathrm{GEP}) = I_0^\mathrm{ESF}(x) + I_{\mathrm{GEP}}^\mathrm{ESF}(x, \varphi_\mathrm{GEP}) 
\end{equation}

with 
\begin{align}
I_0^\mathrm{ESF}(x) &= \left| K_L  e^{\lambda_K}  \right|^2 + \left|  M_L e^{\lambda_M}  \right|^2  
\end{align}

and
\begin{align}
\label{eq2:Idefs}
I_{\mathrm{GEP}}^\mathrm{ESF}(x, \varphi_\mathrm{GEP}) &= 2 \mathrm{Re}  \{   I_{KM}  e^{i  \phi_{KM}}  \} 
\nonumber\\
I_{KM} &= K_L  e^{\lambda_K}  \left( M_L  e^{\lambda_M} \right) ^*
\nonumber\\
\phi_{KM} &= \phi_K - \phi_M = - \varphi_\mathrm{GEP}  +  \frac{1}{1 + \frac{\beta^2}{(\kappa/6)^2}}  \left( 2 k_0 x \right).
\end{align}

The expression for $\phi_{KM}$ is identical to the epxression for $\phi_{CD}$ in equation (\ref{eq:Idefs}). Therefore, the $\tau$-dependent interferences have a phase that depends on $x$ in the same way as in section \ref{sec:fwm}. 

\begin{figure*}[h]
\includegraphics[width=13cm]{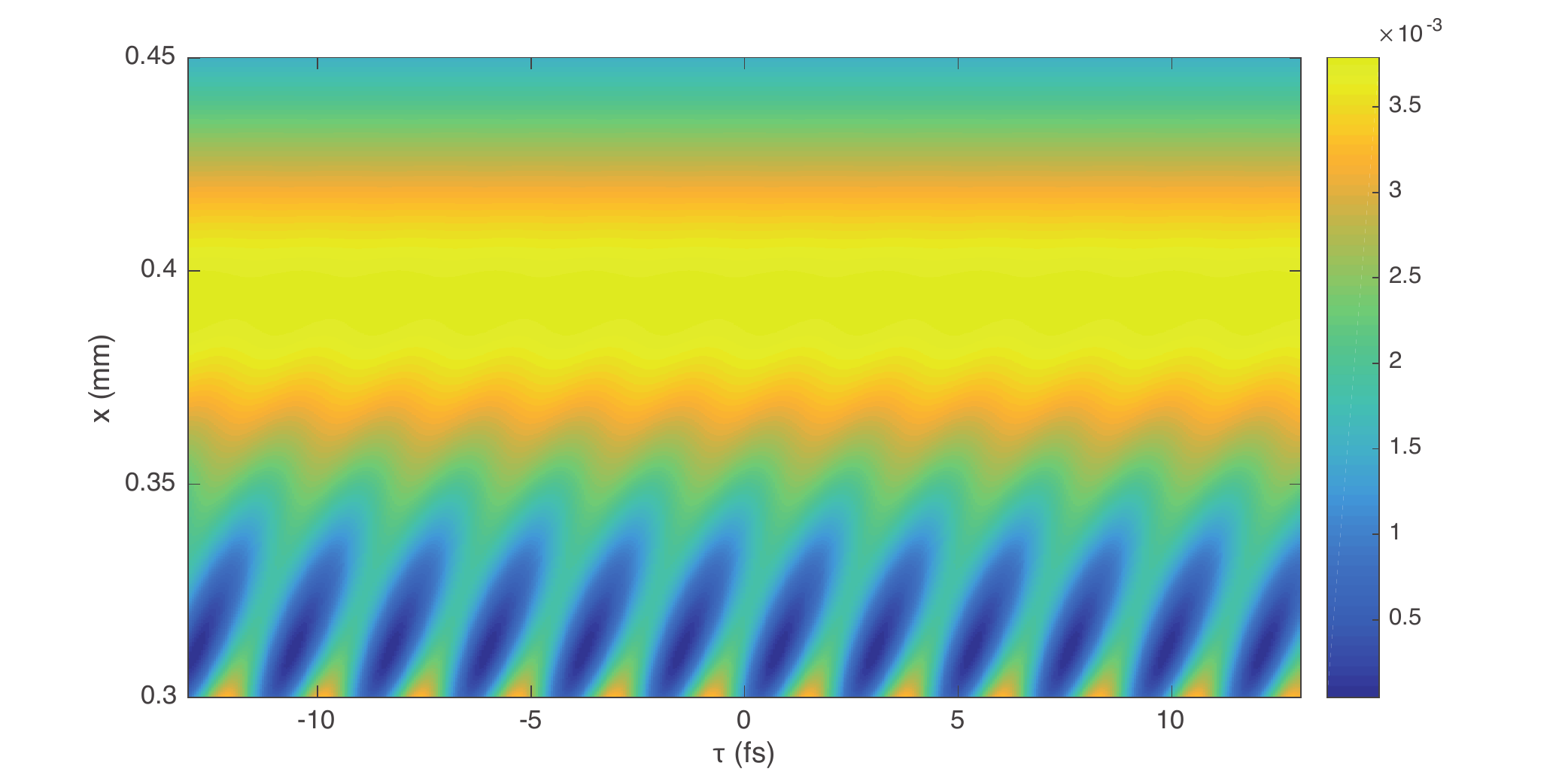}
\caption{\label{fig:ESF} Interferences in ESF according to equation (\ref{eq2:Ixtau}). See Figure \ref{fig:SETUP} for the position of the camera. The same parameter values are used as in Figure \ref{fig:FWM}; additional parameters for ESF are: $b_0 $ = 1,  $b_1$ = 0.1, $b_2$ = 0.1.}
\end{figure*}

As in the case of FWM, the delay scans displayed in Figure \ref{fig:ESF} reveal oscillations with the periodicity of the optical cycle, and the oscillations are most dominant in the region between pulse A and the first diffraction order. Also, the interference fringes are tilted in the $x$-$\tau$-diagram. The intensity of the diffraction order and the strength of the interferences should be reduced for large pulse delays and vanish for no pulse overlap. This is not reproduced by the present model, because the dependence of the coefficients $a_q$ in equation (\ref{eq:grating}), respectively $b_q$ in equation (\ref{eq:eesf}), on the pulse delay $\tau$ is omitted.

\section{\label{sec:ltc} Localized transmission changes}

\subsection{Nonlinear response}

Here it is considered that the medium is not homogeneous in lateral direction ($x$-direction), but rather exhibits localized changes in its transmission. As discussed in Ref. \cite{pati:2016}, the underlying reason for these localized transmission changes could be manifold. One contribution could be localized optical damage in the medium, presumably caused by one of the filaments that are initiated by the laser-induced grating. Another contribution could be processes with high-order nonlinearity, such as nonlinear absorption, which occur most dominantly if a filament forms at the peak of the spatial envelope. 

To investigate localized transmission changes with a simplified model, it is assumed that the electric field after the medium is given by
\begin{equation}
E^{\mathrm{LTC}}(t, x)  =  E(t, x) (1 - T e^{- \frac{1}{2 \xi} x^2} ), 
\end{equation}

where $E(t, x)$ is the field before the medium. This assumption means essentially that the transmission is altered by the factor $1-T$ inside a narrow region with lateral extent $\sqrt{\xi}$. 

To reproduce the interference pattern that is observed in the experiment, the interplay of LTC with FWM and with ESF must be analyzed. Since FWM and ESF yield almost identical interferences (see Figures \ref{fig:FWM} and \ref{fig:ESF}), the following discussion is restricted to the interplay of LTC with FWM. The interplay of LTC and ESF leads essentially to the same result. The electric field after the medium with infinitesimal thickness $l$, altered by both FWM and LTC, is given by
\begin{equation}
\hat E^\mathrm{LTC}(\omega = \omega_0, k) =  \hat E^\mathrm{FWM}(\omega = \omega_0, k)  (1 - T e^{- \frac{1}{2 \xi} x^2} )
\end{equation}
where $\hat E^\mathrm{FWM}$ is defined in equation (\ref{eq:Ew0l}). To avoid very lengthy equations in the following, the approximation
\begin{equation}
\hat E^\mathrm{LTC}(\omega = \omega_0, k) =  \hat E^\mathrm{FWM}(\omega = \omega_0, k)  -  T A_0  e^{-\frac{\eta}{2} (k + k_0)^2}
\end{equation}
with $\eta = \frac{\kappa \xi}{\kappa + \xi}$ is made. The full calculation can be done equivalently and does not significantly alter the appearance of Figures \ref{fig:LTC} and \ref{fig:LTCiso}.

\subsection{Pulse propagation}

The free space propagation from the medium to the camera is calculated similarly as in section \ref{sec:fwmPP}. The electric field at the macroscopic distance $L$ is
\begin{align}
\hat E_L^\mathrm{LTC}(\omega = \omega_0, k) &= \hat E^\mathrm{LTC}(\omega = \omega_0, k)  e^{i \beta k^2}
\nonumber\\
 &= \hat E_L^\mathrm{FWM}(\omega = \omega_0, k)  - T  A_0   e^{-\frac{\eta}{2} (k + k_0)^2} e^{i \beta k^2}.
\end{align}

The inverse Fourier transform yields 
\begin{equation}
\tilde E_L^\mathrm{LTC}(\omega = \omega_0, x) 
=  \tilde E_L^\mathrm{FWM}(\omega = \omega_0, x)  +  H_L  e^{-i \frac{k_0}{\zeta_{\eta}} x}   e^{- \frac{1}{2 \eta \zeta_{\eta}} x^2} 
\end{equation}

with the definitions
\begin{align}
\zeta_{\eta} &= 1 - i \frac{2\beta}{\eta}
\nonumber\\
H_L &= - T A_0 \sqrt{\frac{1}{\zeta_\eta}}   e^{- \frac{\eta}{2} (1 - \frac{1}{\zeta_{\eta}}) k_0^2}.
\end{align}

Corresponding to equation (\ref{eq:group}), the real and imaginary parts in the exponential functions are grouped:

\begin{align}
\tilde E_L^\mathrm{LTC}(\omega = \omega_0, x)   &=  \tilde E_L^\mathrm{FWM}(\omega = \omega_0, x) + H_L  e^{\lambda_H}  e^{i  \phi_H} 
 \nonumber\\
 &=  A_L  e^{\lambda_A}  e^{i  \phi_A} + C_L e^{\lambda_C}  e^{i  \phi_C} +  D_L  e^{\lambda_D}  e^{i  \phi_D} +  H_L  e^{\lambda_H}  e^{i  \phi_H}
\end{align}

with the definitions

\begin{align}
 \lambda_H  &= -\frac{1}{1 + \frac{\beta^2}{(\eta/2)^2}}  \left( \frac{1}{2 \eta} (x - 2 k_0 \beta)^2 - \frac{2 k_0^2 \beta^2}{\eta}  \right)
 \nonumber\\
\phi_H  &=  -\frac{1}{1 + \frac{\beta^2}{(\eta/2)^2}}  \left( \frac{\beta}{\eta^2} x^2 + k_0 x  \right).
\end{align}

\subsection{Interference pattern}

The camera detects the intensity 

\begin{equation}
\label{eq:IG}
I^\mathrm{LTC}(x, \varphi_\mathrm{GEP}) = \left|  \tilde E_L^\mathrm{LTC}(\omega = \omega_0, x)  \right|^2.
\end{equation}

To analyze the interferences, the expression for the intensity is separated into terms that are independent respectively dependent on the GEP:
\begin{equation}
\label{eq:IGxtau}
I^\mathrm{LTC}(x, \varphi_\mathrm{GEP}) = I^\mathrm{LTC}_0(x) + I_{\mathrm{GEP}}^\mathrm{LTC}(x, \varphi_\mathrm{GEP}) .
\end{equation}

\begin{figure*}[h]
\includegraphics[width=13cm]{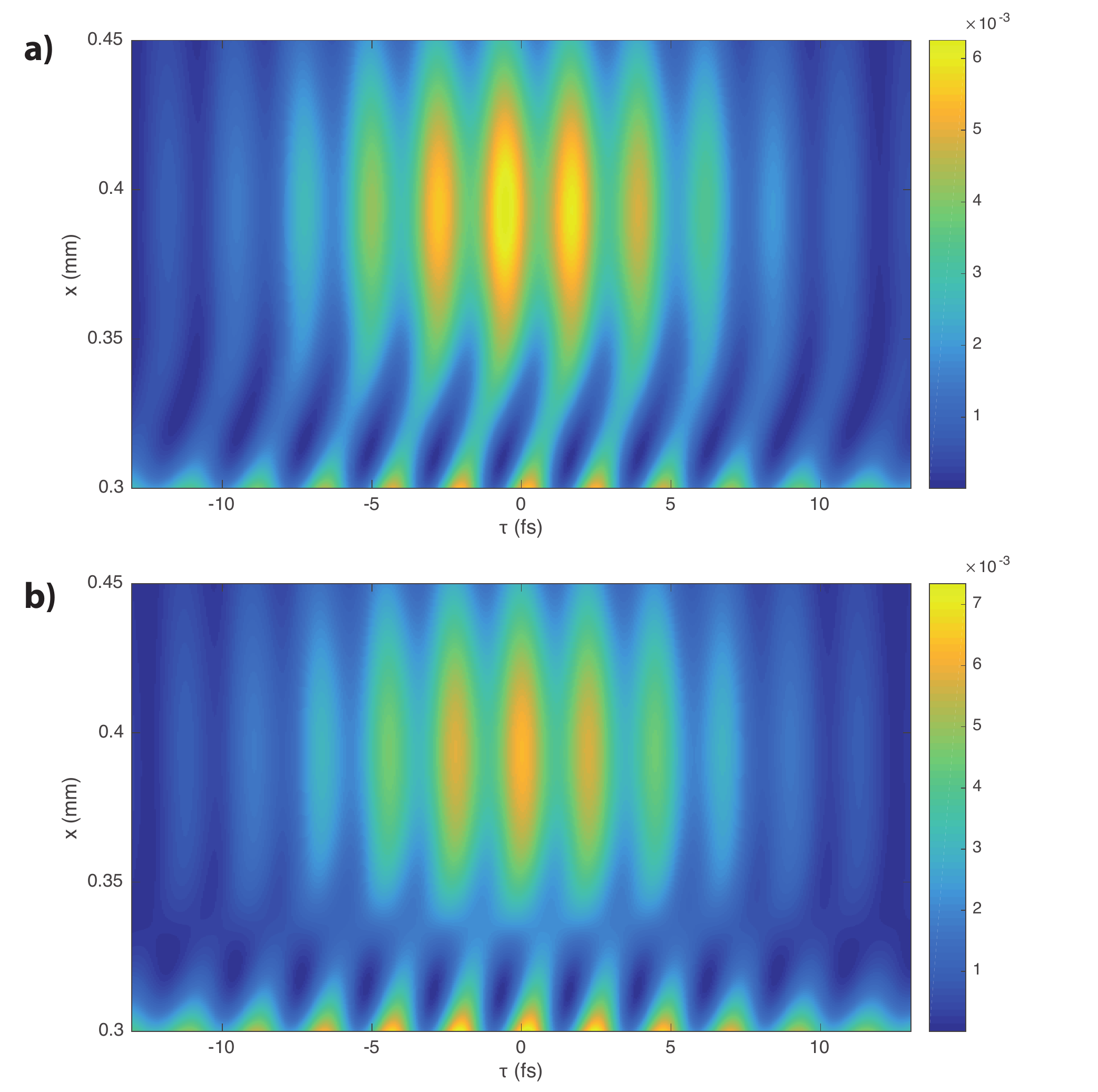}
\caption{\label{fig:LTC} Interferences caused by LTC according to equation (\ref{eq:IGxtau}). See Figure\,\ref{fig:SETUP} for the position of the camera. The same parameter values are used as in Figure \ref{fig:FWM}; additional parameters that determine the LTC are: 
 $T = $ 0.05 for \textbf{a)} respectively $T = $ 0.05$i$ for \textbf{b)} and
$\xi = $3.214e+09\,au (this means that the region where the transmission is altered by the factor $(1-T)$ has a lateral extent of $\sqrt{\xi} = $ 3$\mu$m, which corresponds to the width of the filaments that form in the numerical calculation in Ref. \cite{pati:2016}).}
\end{figure*}

Compared to the analysis in section \ref{sec:fwm}, there is one additional term that is dependent on the GEP:

\begin{align}
\label{eq:IGdefs}
I_{\mathrm{GEP}}^\mathrm{LTC}(x, \varphi_\mathrm{GEP}) &= I_{\mathrm{GEP}}^\mathrm{FWM}(x, \varphi_\mathrm{GEP}) + 2 \mathrm{Re}  \{   I_{HD}  e^{i  \phi_{HD}}  \} 
\nonumber\\
I_{HD} &= H_L  e^{\lambda_H}  \left( D_L e^{\lambda_D} \right) ^*
\nonumber\\
\phi_{HD} &= \phi_H - \phi_D = - \varphi_\mathrm{GEP}  - 
\frac{1}{1 + \frac{\beta^2}{(\eta/2)^2}}  \left( \frac{\beta}{\eta^2} x^2 + k_0 x  \right)  
+\frac{1}{1 + \frac{\beta^2}{(\kappa/6)^2}}  \left(  \frac{9 \beta}{\kappa^2} x^2 + 3 k_0 x \right) .
\end{align}

According to equation (\ref{eq:IGdefs}), $\phi_{HD}$ does also exhibit a dependence on $x$. However, this dependence is negligible with the parameters of Figure \ref{fig:LTC}, so  $\phi_{HD}$ can be approximated by
\begin{align}
\label{eq:ILTCapprox}
\phi_{HD} \approx - \varphi_\mathrm{GEP} .
\end{align}

The delay scans displayed in Figure \ref{fig:LTC} reveal two distinguishable oscillations with the periodicity of the optical cycle. First, there are oscillations in the region between pulse A and the first diffraction order which are tilted in the $x$-$\tau$-diagram. This interference pattern has already been observed in section \ref{sec:fwm} (see Figure \ref{fig:FWM}) and are therefore attributed to four-wave mixing. Second, there are oscillations on the first diffraction order which are vertical in the $x$-$\tau$-diagram. This interference pattern is not present in section \ref{sec:fwm}, therefore it is attributed to localized transmission changes. Both absorption (corresponding to a real parameter $T$, see Figure \ref{fig:LTC} \textbf{a)}) and phase-shift (corresponding to an imaginary parameter $T$, see Figure \ref{fig:LTC} \textbf{b)}) lead to the second kind of interference pattern.

\begin{figure*}[h]
\includegraphics[width=13cm]{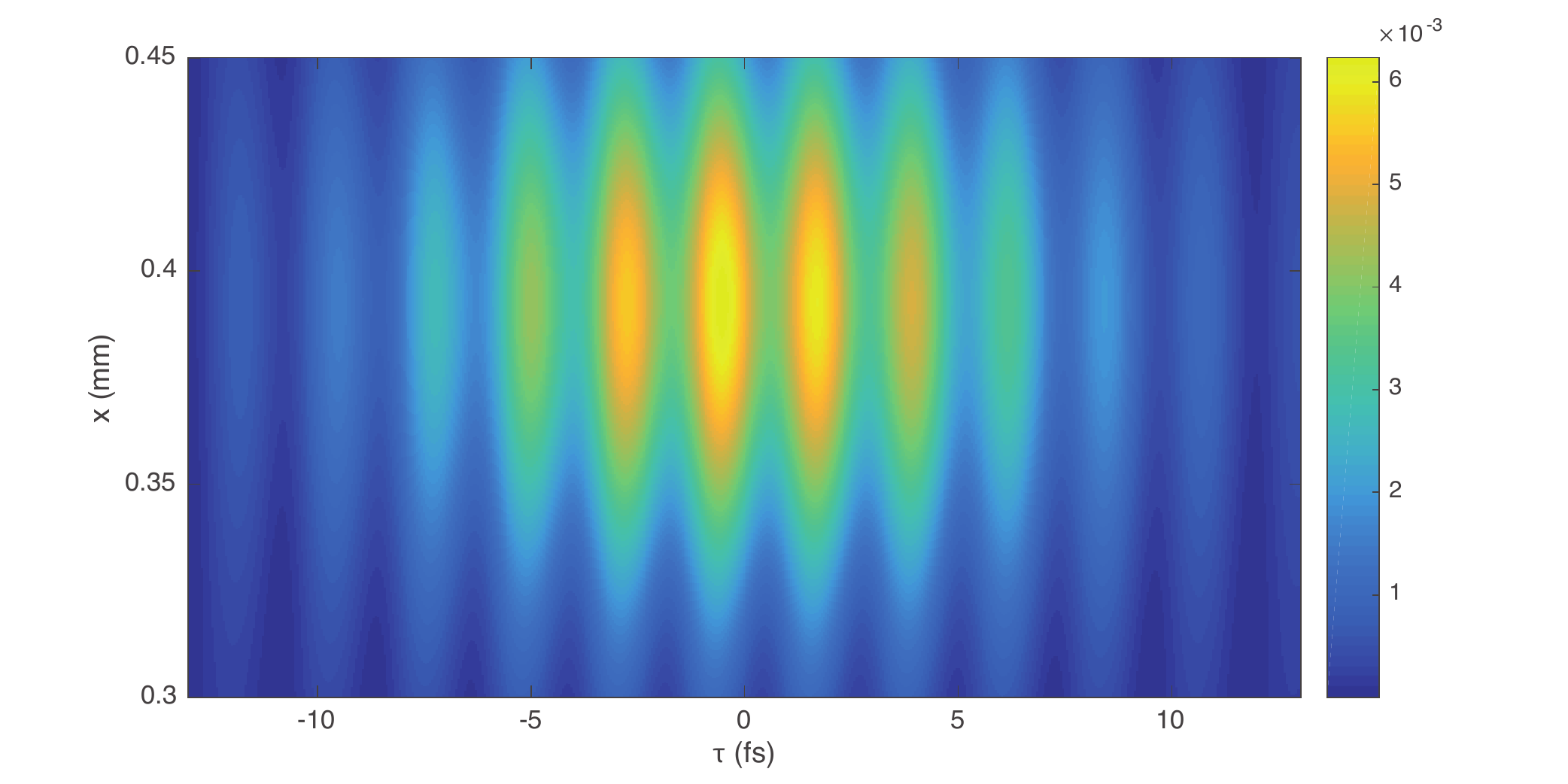}
\caption{\label{fig:LTCiso} Isolated interferences caused by LTC. Displayed is $I^\mathrm{iso}(x)$ according to equation (\ref{eq:ILTC}), without the approximation (\ref{eq:ILTCapprox}), with the same parameter values as in Figure \ref{fig:LTC} \textbf{a)}. See Figure\,\ref{fig:SETUP} for the position of the camera.}
\end{figure*}

The second kind of interference pattern is represented by the term $2 \mathrm{Re}  \{   I_{HD}  e^{i  \phi_{HD}}  \} $ in equation \ref{eq:IGdefs}. To analyze the second kind of interference pattern isolated from the first kind of interference pattern, the quantity 

\begin{equation}
\label{eq:ILTC}
I^\mathrm{iso}(x) = \left| D_L  e^{\lambda_D}  e^{i  \phi_D} +  H_L  e^{\lambda_H}  e^{i  \phi_H}  \right|^2
\end{equation}

is displayed in Figure \ref{fig:LTCiso}. This confirms the observation that the second kind of interference pattern, which is caused by localized changes in transmission, is vertical in the $x$-$\tau$-diagram and most pronounced in the region of the diffraction order. 

It is evident from equation (\ref{eq:IGdefs}) that the slope of the interference fringes in the $x$-$\tau$-diagram depends on $\xi$. The more localized the transmission changes are (the smaller the parameter $\xi$), the weaker is the tilt in the  $x$-$\tau$-diagram. For FIGs. \ref{fig:LTC} - \ref{fig:LTCiso}, the transmission changes are confined to a region of $\sqrt{\xi} = $ 3$\mu$m, and the spacing of the grooves is $2 \pi / (2 k_0) = 39 \mu$m. The calculations show that if the localization of the transmission changes is small compared to the groove spacing, than the interference fringes appear essentially vertical. 

There is an interesting connection in the position-time analogy between GEP and CEP. The time-domain analogue to localization within a groove period is temporal localization within an optical cycle. Processes that are localized within the optical cycle are usually strong-field phenomena, such as strong-field ionization and high-order harmonic generation. These processes also depend on the CEP for laser pulses in the few-cycle regime.

\section{Conclusion}

Laser-induced gratings in the few-groove regime lead to interesting phenomena in nonlinear processes. It was shown that the GEP appears in equations of third-order nonlinear response as the spatial analogue to the temporal CEP. As a consequence, the interferences between orders of self-diffraction can be treated in analogy to the interference between a fundamental beam and a frequency-doubled beam. The first kind of interference pattern, which was experimentally observed in Ref. \cite{pati:2016} and which is dominant in the region between the diffraction orders with tilted interference fringes in the  $x$-$\tau$-diagram, is characteristic for this kind of GEP dependence. Equations have been derived for FWM and ESF yielding this kind of interference pattern. An implication is that the second kind of interference pattern, which was observed only at high intensities close to the damage threshold in Ref. \cite{pati:2016} and which is dominant on the diffraction orders with vertical interference fringes, must be attributed to another mechanism. LTC has been suggested as a mechanism and shown to yield the second kind of interference pattern. 

These findings gain importance in light of the recently growing interest in extreme nonlinear optics in transparent solids. The attribution of readily accessible observations to simple models is important for the understanding and engineering of macroscopic propagation under conditions of extreme nonlinearity.

\bibliographystyle{plain}
\bibliography{2016_Reis}
%\printbibliography
%\printbibliography[heading=bibliography]

\end{document}